\documentclass[12pt]{article}
\title{Local Current Operators for Arbitrary Spin }
\author{W. H. Klink\\
Department of Physics and Astronomy\\
University of Iowa, Iowa City, Iowa}
\begin{document}
\maketitle
\begin{abstract}
Free current operators are constructed for massive particles with arbitrary
spin $j$.  Such current operators are related to representations of the 
$U(N,N)$ type groups and are  covariant under the (extended)
Poincar\'{e} group and charge conjugation, where the charge
conjugation operation is defined as an automorphism on $U(N,N)$
elements.  The currents are also required to satisfy current
conservation, hermiticity, and locality.  The condition that the currents
be local is shown to be equivalent to certain integral constraints on 
form factors.  These constraints are satisfied by writing the currents in
terms of free local spin $j$ fields.  It is shown
that there are
$(2j+1)$ different local currents for a spin $j$ particle, each with an
arbitrary form factor, generalizing the Dirac and
Pauli currents for spin 1/2 particles.  Static properties  of the various
currents are also given.
\end{abstract}
\section{Introduction}
Local current operators are usually defined in terms of bilinear
operators made out of local fields.  If the fields are free fields
written in terms of creation and annihilation operators, the current
operators will consist of a sum of four terms, of the form 
$a^{\dagger}(p,\sigma)a(p^{'},\sigma^{'}),
b^{\dagger}(p,\sigma)b(p^{'},\sigma^{'}),
a^{\dagger}(p,\sigma)b^{\dagger}(p^{'},\sigma^{'})$, and
$a(p,\sigma)b(p^{'},\sigma^{'})$. 
$a(p,\sigma)$ is the annihilation operator for a particle of mass $m$
 $(p\cdot p=m^2)$ and spin $j (-j\le \sigma\le j)$ while
$b(p,\sigma)$ is the annihilation operator for the antiparticle.  As will
be shown, these four bilinears in creation and annihilation operators form
the Lie algebra of the $U(N,N)$ type groups.  The goal of this paper is to
construct current operators out of these Lie algebra elements that have
the correct transformation properties under the extended Poincar\'{e} group,
are conserved, hermitian, local and invariant under charge conjugation. 
To ensure locality, currents will be constructed from free spin $j$ local
fields. It will be shown that there are two classes of currents that give
a total of $2j+1$ independent currents, generalizing the "$\gamma_{\mu}$"
and anomalous currents for spin $1/2$ particles.  Moreover these features
should persist even with arbitrary form factors for the particles.

The construction of currents out of spin $j$ local fields can be
understood in the following way.  Consider a $2(2j+1)$ component field
$\Psi(x)$ made out of two $(2j+1)$ component fields $\phi(x)$ and
$\chi(x)$, each a linear combination of creation and annihilation
operators for particles and antiparticles.  Under a Lorentz
transformation $\Lambda$, $\phi(x)$ goes to
$D^j(\Lambda^{-1})\phi(\Lambda x)$, where $D^j$ is the $(j,0)$
representation of the Lorentz group, while $\chi(x)$ goes to
$\bar{D}^j(\Lambda^{-1})\chi(\Lambda x)$, with 
$\bar{D}^j(\Lambda)=D^j(\Lambda^{-1})^{\dagger}$.  Then $\Psi^{\dagger}
(x)\beta\Psi(x)$ (where $\beta$ is the off diagonal identity matrix
defined following Eq.111) is a Lorentz scalar and a Lorentz tensor can be
defined by making use of the generalized gamma matrices defined by
Weinberg \cite{a}.  These gamma matrices are themselves defined in terms of
generalized sigma matrices, so that
$\Psi^{\dagger}(x)\beta\gamma_{\mu_1...\mu_{2j}}\Psi(x)$ is of the form\\
$\phi^{\dagger}(x)\bar{\sigma}_{\mu_1...\mu_{2j}}\phi(x)+
\chi^{\dagger}(x)\sigma_{\mu_1...\mu_{2j}}\chi(x)$, and is a Lorentz
tensor, invariant under parity and charge conjugation.

There are now two ways to make currents from these tensors; the first
generalizes the way currents are made from spin zero fields, namely by
differentiating the fields to produce a four-vector operator.  Since
there are $2j$ components for the sigma tensors the fields must be
differentiated sufficiently many times to contract with the sigma tensor
indices.  As shown in section 3 there are $j+1$ different ways of doing
this for integer spin, and $j+1/2$ ways for half integer spin, resulting
in $j+1$ different currents for integer spin and $j+1/2$ for half integer
spin with electric, but no magnetic form factors.  The
second  way generalizes the well known spin 1/2 current;  in this case
the fields are again differentiated sufficiently many times to contract
with the sigma matrix tensor indices.  However in this case one index is
left uncontracted, which gives the four-vector current operator. The
number of such currents is now $j$ for integer spin and $j+1/2$ for half
integer spin, resulting in the usual $2j+1$ different possibilities for a
spin $j$ system.

The motivation for constructing such current operators arises from point form
relativistic quantum mechanics \cite{b}, where, for finite degree of freedom
systems, interactions are specified by mass operators which are matrix
elements of vertices, while for infinite degree of freedom systems,
interactions are given by four-momentum operators, obtained by integrating
vertices over the forward hyperboloid.  In both cases the electromagnetic
interaction is formed by coupling the current operators described in this
paper to the photon field, analyzed in detail in the next paper in this
series \cite{c}, to form the electromagnetic vertex for arbitrary spin
particles with arbitrary form factors.  And if these currents are coupled to
a classical field, it becomes possible to investigate some of the so-called
acausal behavior of high spin particles in an external electromagnetic
field \cite{d}.  More generally this series of papers begins an investigation
of many-body relativistic theories in the point form, in which the interacting
four-momentum operator is given solely in terms of creation and annihilation
operators with arbitrary form factors.  

Section 2 reviews properties of representations of the Poincar\'{e} group
needed for the creation and annihilation operators, as well as the
structure of point form quantum mechanics.  The creation and annihilation
operators are then written as bilinears to generate the $U(N,N)$
algebra.  In this context it is shown that the charge and charge
conjugation operators arise naturally from the algebra and its dual.  The
section concludes with a general representation for current operators in
terms of arbitrary form factors that satisfies all the desired properties
(such as charge conservation) except locality;  conversely it is shown
how locality imposes conditions on these form factors.  In the following
sections field theory is used as a guide to construct form factors that
satisfy locality.  In section 3 currents are constructed that generalize
the way currents are made from spin zero fields and in section 4 the
generalization of  $\bar{\Psi}\gamma_\mu\Psi$ for spin 1/2 fields is
given.  In these sections heavy use is made of a paper by Weinberg \cite{a}
on local fields for arbitrary spin particles.  Finally, section 5 shows
how to generalize these results further by including form
factors that keep the currents local.
\section{Local Currents as Representations of $U(N,N)$}
To have a relativistic quantum theory the commutation relations of the
Poincar\'{e} algebra must be satisfied.  In the point form of relativistic
quantum mechanics\cite{b} the commutation relations are to be satisfied by
putting all interactions in the four-momentum operator and leaving the
Lorentz generators kinematic.  The commutation relations can then be
written as
\begin{eqnarray}
[P_{\mu},P_{\nu}]&=&0\\
U_{\Lambda}P_{\mu}U_{\Lambda}^{-1}&=&(\Lambda^{-1})_{\mu}^{\nu} P_{\nu},
\end{eqnarray}
where $U_{\Lambda}$ is the unitary operator representing the Lorentz
transformation $\Lambda$.  Eq.2 emphasizes the fact that since all Lorentz
transformations are kinematic,  it is easier to deal with global Lorentz
transformations than with the infinitesimal generators of the Lorentz
group.  The four-momentum operator $P_{\mu}$ is to be built out of
creation and annihilation operators that themselves transform under
irreducible representations of the Poincare group, in such a way that the
components commute among themselves, which is Eq.1.

If $|p,\sigma>$ is a state of four-momentum $p$ $(p\cdot p=m^2)$ and spin
projection $\sigma$ $(-j\le\sigma\le j)$, then under a Poincar\'{e}
transformation,
\begin{eqnarray}
U_a|p,\sigma>&=&e^{ip\cdot a} |p,\sigma>\\
U_{\Lambda}|p,\sigma>&=&\sum |\Lambda p,\sigma^{'}>D^j_{\sigma^{'}\sigma}
(R_W(v,\Lambda)),
\end{eqnarray}
where $a$ is a four-translation, $D^j_{\sigma^{'}\sigma}()$ is an $SU(2)$
matrix element for spin
$j$ and $R_W(v,\Lambda)$ is a Wigner rotation, an element of the rotation
group $SO(3)$ defined by
\begin{eqnarray}
R_W(v,\Lambda):&=&B^{-1}(\Lambda v)\Lambda B(v).
\end{eqnarray}
$B(v)$ is a boost, a Lorentz transformation satisfying $p=B(v)p^{rest}$,
with $p^{rest}=(m,0,0,0), v=p/m$.  In this paper boosts are always  
canonical spin boosts, defined in the Appendix, Eq.70.

Creation and annihilation operators with the same transformation
properties as states generate multiparticle states from the Fock vacuum:
\begin{eqnarray}
|p,\sigma>&=&a^{\dagger}(p,\sigma)|0>,\\
{[}a(p,\sigma),a^{\dagger}(p^{'},\sigma^{'})]_{\pm}&=&2v_0\delta^3(v-v^{'})
\delta_{\sigma,\sigma^{'}}\\
U_aa(p,\sigma)U_a^{-1}&=&e^{-ip\cdot a}a(p,\sigma)\\
U_\Lambda a(p,\sigma)U_\Lambda^{-1}&=&\sum a(\Lambda p,\sigma^{'})
D^j_{\sigma^{'}\sigma}(R_W(v,\Lambda)^{\ast},
\end{eqnarray}
where the $\pm$ denotes commutator or anticommutator for bosons or
fermions respectively.  Note that the creation and annihilation operators
are normalized in Eq.7 so that they are dimensionless.  With these
creation and annihilation operators the free four-momentum operator can
be written as
\begin{eqnarray}
P_\mu(fr)&=&\sum \int \frac{d^3v}{2v_0}p_\mu a^{\dagger}(p,\sigma)
a(p,\sigma),\nonumber
\end{eqnarray}
and by virtue of the transformation properties of the creation and
annihilation operators, Eq.9, satisfies the point form equations, Eqs.1,2.

As is well known locality requires the existence of antiparticles.  If
$b^{\dagger}(p,\sigma)$ and $b(p,\sigma)$ are the antiparticle creation
and annihilation operators, then these operators will satisfy the same
properties as the particle creation and annihilation operators, Eqs.6
through 9.  The free four-momentum operator now is
\begin{eqnarray}
P_\mu(fr) &=&\sum \int\frac{d^3 v}{2v_0}p_\mu
(a^{\dagger}(p,\sigma)a(p,\sigma)+b^{\dagger}(p,\sigma)b(p,\sigma)),
\end{eqnarray}
and again satisfies the point form equations.

An interacting four-momentum operator is obtained by integrating a local
scalar density over the forward hyperboloid.  For the electromagnetic
interaction this gives
\begin{eqnarray}
P_{\mu}(em)&=&\int d^4 x\delta(x\cdot x-\tau^2)x_\mu J^{\nu}(x)A_{\nu}(x),
\end{eqnarray}
where $J^{\nu}(x)$ is the current operator and $A_{\nu}(x)$ the
local photon field, discussed in reference \cite{c}.  $P_{\mu}(em)$, as
shown in reference \cite{b}, will not satisfy the point form equations
unless the current operator is local.

Free current operators are operators bilinear in the four creation and
annihilation operators.  If the indices $(p,\sigma)$ are replaced by $i$,
the commutation relations of these bilinears can be written as 
\begin{eqnarray}
[a^{\dagger}(i)a(j),b^{\dagger}(k)b(l)]&=&0\\
{[}a^{\dagger}(i)a(j),a^{\dagger}(k)b^{\dagger}(l)]&=&
\delta_{j,k}a^{\dagger}(i)b^{\dagger}(l)\\
{[}a^{\dagger}(i)a(j),a(k)b(l)]&=&-\delta_{i,k}a(j)b(l)\\
{[}a(i)b(j),a^{\dagger}(k)b^{\dagger}(l)]&=&\delta_{i,k}\delta_{j,l}
\pm\delta_{j,l} a^{\dagger}(k)a(i)\pm\delta_{i,k} b^{\dagger}(l)b(j).
\end{eqnarray}  
It should be noted that only the last commutation relation, Eq.15
distinguishes between bosons and fermions.  The commutation relations
given above are those of the Lie algebra of the group $U(N,N)$, defined by
\begin{eqnarray}
U(N,N):&=&[g\in GL(2N,C)|g\tilde{\beta} g^{\dagger}=\tilde{\beta}],\\
\tilde{\beta}&=&diag(I,-I),
\end{eqnarray}
where $I$ is the N dimensional identity matrix.

The charge operator is defined as $Q:=e\sum(a^{\dagger}(i)a(i)-
b^{\dagger}(i)b(i))$ and the $U(1)$ group that it generates is
dual to $U(N,N)$ in the sense defined in reference \cite{e}.  More
generally,
$U(N,N)$ is dual to $U(n)$ in that, on Fock space, the irreducible
representations of $U(n)$ fix those of $U(N,N)$ and vice versa.  Though
internal symmetries are not considered in this paper, the theory of dual
representations provides a natural setting for the $U(n)$ internal
symmetries, a topic that will be taken up in a later paper.  Also the
charge conjugation operation is the element in the automorphism group of
$U(N,N)$ given by $c=\beta$, where $\beta$ is the off diagonal identity
given following Eq.111 with $I$ the same identity matrix as in Eq.17.  It
is easily checked that if
$g$ is an element of $U(N,N)$, then so is $cgc$;  thus $c$ maps the
$U(N,N)$ Lie algebra into itself.

The discrete, infinite dimensional, unitary representations of the Lie
algebra of $U(N,N)$ are specified by $a(i)b(j)|0>=0$.  In
this paper $|0>$ will be taken to be the Fock vacuum, defined by the
stronger condition, $a(i)|0>=b(j)|0>=0$, for all $(i,j)$.  There are
however many other discrete series irreducible representations of $U(N,N)$
given for example, by the "vacuum" $|0>_n=a^{\dagger}(i_1)...a^{\dagger}
(i_n)|0>$, for which the charge of the "vacuum" is not zero.

Returning to particle labels, free current operators are now defined as
\begin{eqnarray}
J_\mu(0)&=&\sum\int\frac{d^3 v_1}{2v^0_1}\frac{d^3 v_2}{2v^0_2}(F_\mu^a
(v_1\sigma_1,v_2\sigma_2)a^{\dagger}(p_1\sigma_1)a(p_2\sigma_2)
\nonumber\\
&&+F_\mu^b(v_1\sigma_1,v_2\sigma_2)b^{\dagger}(p_1\sigma_1)b(p_2\sigma_2)
\nonumber\\
&&+F_\mu^c(v_1\sigma_1,v_2\sigma_2)a^{\dagger}(p_1\sigma_1)
b^{\dagger}(p_2\sigma_2)\nonumber\\
&&+F_\mu^d(v_1\sigma_1,v_2\sigma_2)a(p_1\sigma_1)b(p_2\sigma_2))\\
&=&\sum\int dv_1
dv_2(F_{\mu}^a(1,2)a^{\dagger}(1)a(2)+F_{\mu}^b(1,2)b^{\dagger}(1)b(2)
\nonumber\\
&&+F_{\mu}^c(1,2)a^{\dagger}(1)b^{\dagger}(2)+F_{\mu}^d(1,2)a(1)b(2),\
\end{eqnarray}
and the goal is to choose the $F_\mu(1,2)$ in such a way that the current
operator is Poincar\'{e} covariant, conserved, local, hermitian, and
invariant under charge conjugation.

Translational covariance is achieved by defining $J_\mu(x):=U_xJ_\mu(0)
U_x^{-1}$, where $U_x$ is defined in Eq.8.  Lorentz covariance, charge
conservation, parity and time-reversal covariance lead to conditions on
the $F$'s discussed in reference \cite{f}.  In particular charge
conservation is $[P^{\mu}(fr),J_\mu(0)]=0$. Hermiticity leads
to the following relations on the $F's$:
$J_\mu(0)^\dagger=J_\mu(0)$ means that $F_\mu^a(2,1)^{\ast}=F_\mu^a(1,2)$,
$F_\mu^b(2,1)^{\ast}=F_\mu^b(1,2)$, and
$F_\mu^d(1,2)=F_\mu^c(1,2)^{\ast}$.

Charge conjugation invariance is a bit more  complicated.  Following
Weinberg \cite{a}, the unitary operator representing charge conjugation
operates on the creation and annihilation operators in the following way:
\begin{eqnarray}
U_c a(1)U_c^{-1}&=&\eta_c b(1)\\
U_c b(1)U_c^{-1}&=&{\eta}_c ^{\ast} a(1),
\end{eqnarray}
with $\eta_c\eta_c^{\ast}=1$.  Further, $U_cQU_c^{-1}=-Q$, where 
$Q:=\int d^3 x J_{\mu=0}(\vec{x})$ is the charge operator, which implies
that charge conjugation anticommutes with the current operator.  Putting
these conditions together then gives the following conditions on the
$F's$:  $F_\mu^b(1,2)=-F_\mu^a(1,2)$ and $F_\mu^c(2,1)=-F_\mu^c(1,2)$.

The current operator can thus be written as
\begin{eqnarray}
J_\mu(0)&=&\sum\int
dv_1dv_2[F_\mu^a(1,2)(a^{\dagger}(1)a(2)-b^{\dagger}(1)b(2))\nonumber\\
&&+F_\mu^c(1,2)a^{\dagger}(1)b^{\dagger}(2)+F_\mu^c(1,2)^{\ast}a(1)b(2)],
\end{eqnarray}
and the charge operator formed from this $J_\mu(0)$ agrees with the
charge operator given following Eq.17. 

 Now the $F$'s are one-particle matrix
elements of the current operator;  for example
\begin{eqnarray}
F_\mu^a(v_1\sigma_1,v_2\sigma_2)&=&<p_1\sigma_1|J_\mu(0)|p_2\sigma_2>
\nonumber\\
&=&<0|a(p_1\sigma_1)J_\mu(0)a^{\dagger}(p_2\sigma_2)|0>.
\end{eqnarray}
By exploiting the Poincar\'{e} tensor transformation properties of the
current operator, reference \cite{f} shows that the matrix element, Eq.23,
can be written as a covariant (Clebsch-Gordan coefficient of the Poincare
group) times an invariant (form factor).  A simple way of deriving these
results is to choose a standard frame (the Breit frame) and use
invariance under z axis rotations, parity, and time reversal to get
conditions on the standard (invariant form factor) matrix elements:
\begin{eqnarray}
F_\mu^a(Q^2)_{r_1r_2}:&=&<p_1(st)r_1|J_\mu(0)|p_2(st)r_2>\\
F_{\mu=0}^a(Q^2)_{r_1r_2}&=&<p_1(st)r_1|U_{R_z}^{-1}
U_{R_z}J_{\mu=0}(0)U^{-1}_{R_z}U_{R_z}|p_2(st)r_2>\nonumber\\
&=&A_{r_1}(Q^2)\delta_{r_1r_2};\nonumber\\
<\frac{\vec{Q}}{2},r_1|J_{\mu=0}|-\frac{\vec{Q}}{2},r_2>
&=&G_C(Q^2)\sigma_{0...0}+G_Q(Q^2)\sigma_{ij0...0}\frac{Q_i}{2m}
\frac{Q_j}{2m}+...
\end{eqnarray}
where $A_{r_1}$ is diagonal as a matrix in the $r_1,r_2$ variables and the
$r_1,r_2$ are invariant spin projection variables, ranging between $-j\le
r_1,r_2\le j$.  The standard vectors can be written in several ways, all
depending on the invariant momentum transfer only:
$p_1(st)=(E(st),0,0,Q/2)=m(ch\alpha,0,0,sh\alpha),\\ p_2(st)=(E(st),
0,0,-Q/2)=m(ch\alpha,0,0,-sh\alpha)$ where\\
$q^2=(p_1(st)-p_2(st))^2=-4m^2sh^2\alpha=-Q^2$ is the invariant momentum
transfer. 
$E(st)=\sqrt{m^2+(Q/2)^2}=mch\alpha$ and $R_z$ is a rotation about the z
axis through an angle $\phi$.  Parity and time-reversal further restrict
the
$F^a_{\mu=0}$ as shown in reference \cite{f}. 

It is possible to further decompose the unknown invariants $A_{r_1}(Q^2)$
into various moments, such as charge and quadrupole moments, by writing
the spin indices in terms of powers of spin matrices.  It is then natural
to use the sigma tensors introduced by Weinberg \cite{a};  the
sigma tensors are themselves powers of spin matrices, with the property
that for a particle of spin $j$ there are
$(2j+1)^2$ tensors. In
 Eq.25 the Breit frame matrix element is written in terms of charge
($G_C(Q^2))$, quadrupole ($G_Q(Q^2))$ and higher multipole terms times
sigma matrices defined in the appendix,  Eq.82;  thus the sigma matrices
provide a natural representation for spin indexed current matrix elements,
which can be used to compare with the form factors obtained in sections 3
and 4.  At momentum transfer $Q^2=0$ the various moments give the static
properties of particles, such as their charge.

 A similar analysis shows that the space
components of the Breit frame current matrix elements are of the form
\begin{eqnarray}
<\frac{\vec{Q}}{2},r_1|J_{\mu=i}(0)|-\frac{\vec{Q}}{2},r_2>&=&
i\epsilon_{ijk}\frac{Q_j}{2m}[G_M(Q^2)\sigma_{k0...o}\nonumber\\
&&+G_{M_2}(Q^2)\sigma_{klm0...0}\frac{Q_l}{2m}\frac{Q_m}{2m}+...
\end{eqnarray}
where $G_M(Q^2)$ is the magnetic form factor and is the first in the
series of higher magnetic form factors.  Note that  
$\vec{Q}$ dotted into the Breit frame matrix element is zero,
which is charge conservation.

To get the general form of the one-particle matrix element, Eq.23, let
$\Lambda(v_1v_2)$ be the Lorentz transformation taking $p_1(st)$ to $p_1$
and $p_2(st)$ to $p_2$.  Such a Lorentz transformation is a coset
representative of the Lorentz group $SO(1,3)$ with respect to the
subgroup $SO(2)$ of z axis rotations, and is specified by five parameters.
It can be written as 
\begin{eqnarray}
\Lambda(v_1v_2)&=&B(v)R(\hat{n})\\
&=&(\frac{p_1+p_2}{2E(st)},w_1,w_2,\frac{p_1-p_2}{\sqrt{Q^2}}),
\end{eqnarray}
where $v=\frac{p_1+p_2}{2E(st)}$, the rotation $R(\hat{n})$ takes the unit
vector along the z direction to the unit vector $\hat{n}$ formed from
$B^{-1}(v) (p_1-p_2)$, and the two length minus one space-like vectors
$w_i$ are fixed by Eq.27.

Inserting $U^{-1}_{\Lambda(v_1v_2)}U_{\Lambda(v_1v_2)}$ before and after
$J_\mu(0)$ in Eq.24, and making use of the Lorentz transformation
properties of single particle states, Eq.4,  then gives the final result
for $F^a_{\mu}(v_1\sigma_1,v_2\sigma_2)$ in terms of covariants times
invariants:
\begin{eqnarray}
F_\mu^a(v_1\sigma_1,v_2\sigma_2)&=&\sum\Lambda_\mu^{\nu}(v_1v_2)D^j_
{\sigma_1r_1}(R_W(v_1(st),\Lambda(v_1v_2)))\nonumber\\
&&F^a_\nu(Q^2)_{r_1r_2}D^j_{r_2\sigma_2}(R_W(v_2(st),\Lambda(v_1v_2))^{-1});
\end{eqnarray}
the Wigner rotations appearing in Eq.29 will be worked out in detail in
the following sections.  Note that if $v_1$ and $v_2$ are chosen in their
standard (Breit) frame, $\Lambda(v_1v_2)$ is the identity element, as are
the arguments of the two $D^j$ functions.  Charge conservation means 
$(p_1-p_2)^{\mu}F^a_{\mu}(1,2)=0$ and is a consequence of the fact that
$(p_1-p_2)^{\mu}\Lambda_{\mu}^{\nu}=0$ for $\nu=0,1,2$; for $\nu=3,$ the
invariant form factor is zero (see Eq.26).

A similar analysis holds for
$F_{\mu}^c(v_1\sigma_1,v_2\sigma_2)$;  that is
\begin{eqnarray}
F_\mu^c(v_1\sigma_1,v_2\sigma_2)&=&<p_1\sigma_1,p_2\sigma_2|J_\mu(0)|0>\\
&=&\sum\Lambda_{\mu}^{\nu}(v_1v_2)D_{\sigma_1
r_1}^j(R_W(v_1(st),\Lambda(v_1v_2)))\nonumber\\
&&D_{\sigma_2
r_2}^j(R_W(v_2(st),\Lambda(v_1v_2)))F^c_{\nu}(Q^2)_{r_1r_2}\\
&=&\sum \Lambda_{\mu}^{\nu}(v_1v_2)D_{\sigma_1
r_1}^j(R_W(v_1(st),\Lambda(v_1v_2)))\nonumber\\
&&F_{\nu}^c(Q^2)_{r_1r_2}
D_{-r_2\sigma_2}^j(R_W(v_2(st),\Lambda(v_1v_2))^{-1})^{\ast},\
\end{eqnarray}
with $F_{\mu=0}^c(Q^2)_{r_1r_2}$ antidiagonal in $r_1$ and $r_2$.
The goal now is to relate $F^a_\mu(1,2)$ and $F^c_\mu(1,2)$ through
locality.

Locality means that the commutator of the current with itself should
vanish for space-like separation:
\begin{eqnarray}
{[}J_\mu(x),J_\nu(0)]&=&0
\end{eqnarray}
for $x\cdot x<0$.  Though $J_\mu(0)$ is not an operator in that it
takes elements of the Fock space out of the Fock space, its matrix
elements, $(\Psi, J_\mu(0)\Psi)$, for elements $\Psi$  in the Fock space
are well defined.  But in general products of such operators are not
well-defined.  Here however the underlying $U(N,N)$ structure guarantees
that the commutators are well-defined, and so the problem to be solved is
when the commutator is zero for $x$ spacelike.  Using Eq.22 and
translating by $x$ using Eq.8 makes it possible to calculate the
commutator, $[J_\mu(x),J_\nu(0)]$;  the condition that the commutator be
zero for $x$ spacelike then reduces to the following conditions on
the $F's$:
\begin{eqnarray}
\sum\int dv_3[F_\mu^a(1,3)F_\nu^a(3,2)e^{-ip_3\cdot x}+F_\mu^c(1,3)
F_\nu^c(3,2)^{\ast}e^{ip_3\cdot x}]&=&0\\
\sum\int dv_3[F_\mu^a(1,3)F_\nu^c(3,2)e^{-ip_3\cdot x}+F_\mu^c(1,3)
F_\nu^a(3,2)e^{ip_3\cdot x}]&=&0.\\
\sum\int dv_3 dv_4[F_\mu^c(3,4)^{\ast}
F_\nu^c(4,3)e^{-i(p_3+p_4)\cdot x}\nonumber\\+
F_\mu^c(3,4)F_\nu^c(4,3)^{\ast}e^{i(p_3+p_4)\cdot x}]&=&0
\end{eqnarray}
The question that now must be dealt with is how to choose $F_{\mu}^a$ and
$F_{\mu}^c$ so that these constraints are satisfied.  The
next three sections use field theory as a guide to answer this question.

\section{Local Currents with Electric Form Factors}
The simplest way to generate local currents from spin $j$ fields is
to generalize the way currents are obtained from spin zero fields.   That
is, the  current operator $J_{\mu}(x)=i(\phi^{\dagger}(x)\frac{\partial}
{\partial
x^{\mu}}\phi(x)-(\frac{\partial}{\partial x^{\mu}}\phi^{\dagger}(x))
\phi(x))$, where $\phi(x)$ is the field defined in Eq.99 for $j=0$, can
be generalized to
\begin{eqnarray}
J_{\mu}(x)&=&i(\Psi^{\dagger}(x)\beta\frac{\partial}{\partial x^{\mu}}
\Psi(x)-(\frac{\partial}{\partial x^{\mu}}\Psi^{\dagger}(x))\beta\Psi(x)),
\end{eqnarray}
where $\Psi(x)$ is the $2(2j+1)$ component field defined in Eq.110.
However, such a current will have only a charge, and no other moments; 
the goal of this section is to further generalize the current in Eq.37 so
that it also has higher electric moments.  Then in the next section 
currents that also have magnetic moments are analyzed.

The idea is to use higher derivatives in conjunction with the 
gamma tensors defined in the appendix, Eq.111  to
form more complex local current operators out of local fields.  To that
end define the current operator as
\begin{eqnarray}
J_{\mu}(x):&=&i((\frac{\partial}{\partial
x_{\mu_1}}...\frac{\partial}{\partial
x_{\mu_k}})\Psi^{\dagger}(x))\beta\gamma_{\mu_1...\mu_{2j}}(\frac{\partial}{\partial
x_{\mu_{k+1}}}...\frac{\partial}{\partial
x_{\mu_{2j}}})\frac{\partial}{\partial x^{\mu}}\Psi(x)\
\end{eqnarray}
plus three other terms that are needed to guarantee hermiticity and
charge conservation; such a current is conserved by virtue of the fields
satisfying the Klein-Gordan equation.  It also reduces to Eq.37 if
$k=0$ by virtue of the generalied Dirac equation, Eq.109.  $\beta$ is the
generalization of the four dimensional matrix for spin 1/2 and is defined
after Eq.111.

Since the free fields defined in Eqs.99,106 are sums of creation and
annihilation operators, the four terms in the current operator, Eq.38
will be of the desired form, Eq.22;  moreover, the $F$'s can then be
obtained by taking matrix elements of the current operator:
\begin{eqnarray}
F^a_\mu(1,2)&=&<0|a(1)J_{\mu}(0)a^{\dagger}(2)|0>\\
&=&(-1)^{k+1}(p_1+p_2)_{\mu}[D^{j}(B(v_1)^{\dagger}\bar{\sigma}_{\mu_1...\mu_k\mu_{k+1}...
\mu_{2j}}D^j(B(v_2))\nonumber\\
&&+\bar{D}^{j}(B(v_1))^{\dagger}\sigma_{\mu_1...\mu_k\mu_{k+1}...\mu_{2j}}
\bar{D}^j(B(v_2))]\nonumber\\
&&v_1^{\mu_1}...v_1^{\mu_k}v_2^{\mu_{k+1}}...v_2^{\mu_{2j}}+hc
\nonumber\\
&=&(-1)^{k+1}(p_1+p_2)_{\mu}[D^j(B(v_1))\bar{\sigma}_k(1,2)
D^j(B(v_2))\nonumber\\
&&+\bar{D}^j(B(v_1))\sigma_{k}(1,2)\bar{D}^j(B(v_2))],
\end{eqnarray}
where
$\sigma_k(1,2):=\sigma_{\mu_1...\mu_k\mu_{k+1}...\mu_{2j}}(v_1^{\mu_1}
...v_1^{\mu_k}v_2^{\mu_{k+1}}...v_2^{\mu_{2j}}+v_2^{\mu_1}...v_2^{\mu_k}
v_1^{\mu_{k+1}}...v_1^{\mu_{2j}})$ and is symmetric under the
interchange of 1 and 2.  To show that the matrix element $F^a_{\mu}(1,2)$
has the general form given by Eq.29, that part of Eq.40 in
square brackets is designated as A and manipulated in the following way:
\begin{eqnarray}
A&=&D^j(B(v_1))\bar{\sigma}_k(1,2)D^j(B(v_2))+\bar{D}^j(B(v_1))
\sigma_k(1,2)\bar{D}^j(B(v_2))\\
&=&D^j(B^{-1}(v_1))[D^j(B(v_1)B(v_1))\bar{\sigma}_k(1,2)\nonumber\\
&&+\sigma_k(1,2)D^j(B^{-1}(v_2)B^{-1}(v_2))]D^j(B(v_2))\nonumber\\
&=&D^j(B^{-1}(v_1)\Lambda(1,2))[D^j(\Lambda^{-1}(1,2)D^j(B(v_1)B(v_1))
D^{j}(\Lambda^{-1}(1,2))^{\dagger}\nonumber\\
&&D^{j}(\Lambda(1,2))^{\dagger}\bar{\sigma}_k(1,2)
D^j(\Lambda(1,2)+D^j(\Lambda^{-1}(1,2)
\sigma_k(v_1v_2)D^{j}(\Lambda^{-1}(1,2))^{\dagger}\nonumber\\
&&D^{j}(\Lambda(1,2))^{\dagger}D^j(B^{-1}(v_2)B^{-1}(v_2))D^j(\Lambda(1,2))]
D^j(\Lambda^{-1}(1,2)B(v_2))\nonumber\\
&=&D^j(R_W(v_1(st),\Lambda(1,2))[\bar{\sigma}_k(v_1(st)v_2(st))\nonumber\\
&&+\sigma_k(v_1(st)v_2(st))]D^j(R_W^{-1}(v_2(st),\Lambda(1,2))),
\end{eqnarray}
which is of the form given in Eq.29 when $F_{\nu=i}^a(Q^2)$  is zero; 
that is, there are only electric and no magnetic form factors, as
predicted.  If the matrix element, Eq.40 is evaluated in the standard
(Breit) frame, the two Wigner rotations are the identity and what remains
is exactly the invariant electric form factor, namely
\begin{eqnarray}
F_{\mu=0}^a(Q^2)_{r_1r_2}
&=&(-1)^{k+1}(\sigma_k(v_1(st)v_2(st))+\bar{\sigma}_k
(v_1(st)v_2(st)))_{r_1r_2}.
\end{eqnarray}
It is of course possible to take linear combinations of these invariant
form factors, summing over the k variable, which then gives the most
general class of electric form factors.

Similarly, the matrix element $F_{\mu}^c(1,2)$ is given by
\begin{eqnarray}
F_{\mu}^c(1,2)&=&<p_1\sigma_1,p_2\sigma_2|J_\mu(0)|0>\nonumber\\
&=&(-p_1+p_2)_{\mu}[(-1)^{2j}D(B(v_1))\bar{\sigma}_k(1,2)D(B(v_2))
\nonumber\\
&&+\bar{D}(B(v_1))\sigma_k(1,2)\bar{D}(B(v_2))]C^{-1},
\end{eqnarray}
and can also be manipulated into the form of Eq.32. C is the conjugation
matrix defined after Eq.102 If the expressions for $F_{\mu}^a$ in Eq.40 and
$F_{\mu}^c$ in Eq.44 are substituted into Eqs.34,35,36 which expresses the
locality of the current operator, it is straightforward but tedious to
show that locality is satisfied;  this of course is not surprising, since
these matrix elements came from currents that were given in terms of local
fields.

Finally, some examples of invariant form factors  for spin 1/2, 1, and
3/2 are given by the following expressions:
\begin{eqnarray}
F_{\mu=0}^a(Q^2)&=&\sigma_k(1,2)+\bar{\sigma}_k(1,2)\\
&=&(\sigma_{\mu_1,...\mu_k \mu_{k+1}...\mu_{2j}}+\bar{\sigma}
_{\mu_1...\mu_k\mu_{k+1}...\mu_{2j}})\nonumber\\
&&(v_1^{\mu_1}(st)...v_1^{\mu_k}(st)v_2^{\mu_{k+1}}(st)...v_2^{\mu_{2j}}(st)
\nonumber\\
&&+v_2^{\mu_1}(st)...v_2^{\mu_k}(st)v_1^{\mu_{k+1}}(st)...v_1^{\mu_{2j}}
(st))\\
j=1/2:F_{\mu=0}^a(Q^2)&=&(\sigma_{\mu}+\bar{\sigma}_\mu)(v_1^{\mu}(st)+
v_2^{\mu}(st))\nonumber\\
&=&(\sigma_0+\bar{\sigma}_0)2ch\alpha\nonumber\\
&=&4ch\alpha I;\\
j=1:F_{\mu=0}^a(Q^2)&=&(\sigma_{\mu_1\mu_2}+\bar{\sigma}_{\mu_1\mu_2})
(v_1^{\mu_1}(st)v_1^{\mu_2}(st)+v_2^{\mu_1}(st)v_2^{\mu_2}(st))\nonumber\\
&=&2\sigma_{00}(2ch^2\alpha)+2\sigma_{33}(2sh^2\alpha)\nonumber\\
&=&4\left[\begin{array}{ccc}ch2\alpha&&\\&1&\\&&ch2\alpha\end{array}
\right],(k=2j=2);\\
j=1:F_{\mu=0}^a(Q^2)&=&(\sigma_{\mu_1\mu_2}+\bar{\sigma}_{\mu_1\mu_2})
(v_1^{\mu_1}(st)v_2^{\mu_2}(st))\nonumber\\
&=&2\sigma_{00}ch^2\alpha-2\sigma_{33}sh^2\alpha\nonumber\\
&=&2ch^2\alpha I-2(2S_z^2-I)sh^2\alpha\nonumber\\
&=&2\left[\begin{array}{ccc}1&&\\&ch2\alpha&\\&&1\end{array}\right],(k=1);\\
j=3/2:F_{\mu=0}^a(Q^2)&=&(\sigma_{\mu_1\mu_2\mu_3}+
\bar{\sigma}_{\mu_1\mu_2\mu_3})(v_1^{\mu_1}(st)v_1^{\mu_2}(st)v_1^{\mu_3}(st)
+1<->2)\nonumber\\
&=&4\sigma_{000}ch^3\alpha+4\sigma_{330}sh^2\alpha ch\alpha\nonumber\\
&=&4\left[\begin{array}{cccc}ch3\alpha&&&\\&ch\alpha&&\\
&&ch\alpha&\\&&&ch3\alpha\end{array}\right],(k=2j=3);\\
j=3/2:F_{\mu=0}^a(Q^2)&=&(\sigma_{\mu_1\mu_2\mu_3}+
\bar{\sigma}_{\mu_1\mu_2\mu_3})\nonumber\\
&&(v_1^{\mu_1}(st)v_1^{\mu_2}(st)v_2^{\mu_3}(st)
+v_1^{\mu_1}(st)v_2^{\mu_2}(st)v_2^{\mu_3}(st))\nonumber\\
&=&4\sigma_{000}ch^3\alpha-4\sigma_{330}sh^2\alpha ch\alpha\nonumber\\
&=&4ch\alpha\left[\begin{array}{cccc}1&&&\\&1+\frac{2}{3}sh^2\alpha&&\\
&&1+\frac{2}{3}sh^2\alpha&\\&&&1\end{array}\right],(k=2).
\end{eqnarray}
As discussed after Eq.25 the momentum transfer squared is given by
$\frac{Q^2}{4m^2}=sh^2\alpha$.  For each value of $j$, $k$ goes from $0$
to $2j$; but since $\sigma_k(v_1v_2)$ is symmetric under the interchange
of
$1$ and
$2$, there are $j+1$ different possibilities for integer spin and $j+1/2$
possibilities for half integer spin.  For example, there are two different
electric form factors for both $j=1$ and $j=3/2$, as seen in Eqs.48
through 51.  
\section{Local Currents with Electric and Magnetic Form Factors}
The goal of this section is to construct local currents that generalize
the well-known $\bar{\Psi}\gamma_{\mu}\Psi$ construction for spin 1/2. 
To that end use is again made of the generalized gamma matrices defined
in the appendix, Eq.111, to define the following current operator for
arbitrary spin:
\begin{eqnarray}
J_\mu(x):&=&(\frac{\partial}{\partial
x_{\mu_1}}...\frac{\partial}{\partial x_{\mu_k}})
\Psi^{\dagger}(x)\beta\gamma_{\mu\mu_1...\mu_k\mu_{k+1}...\mu_{2j-1}}
\nonumber\\&&(\frac{\partial}{\partial x_{\mu_{k+1}}}...
\frac{\partial}{\partial x_{\mu_{2j-1}}})\Psi(x)+...
\end{eqnarray}
with the extra terms needed to satisfy current conservation and
hermiticity.  These terms will be added on after the expression in Eq.52
has been decomposed into the appropriate matrix elements.  As in section
3 matrix elements of Eq.52 are evaluated at $x=0$:
\begin{eqnarray}
F_\mu^a(1,2)&=&<p_1\sigma_1|J_\mu(0)|p_2\sigma_2>\\
&=&(-1)^{k+1}(D^j(B(v_1))\bar{\sigma}_k(1,2)_\mu D^j(B(v_2))\nonumber\\
&&+\bar{D}^j(B(v_1))\sigma_k(1,2)_\mu\bar{D}^j(B(v_2))),
\end{eqnarray}
where
now $\sigma_k(1,2)_\mu:=\sigma_{\mu\mu_1...\mu_k\mu_{k+1}...\mu_{2j-1}}
(v_1^{\mu_1}...v_1^{\mu_k}v_2^{\mu_{k+1}}...v_2^{\mu_{2j-1}}\\
+v_2^{\mu_1}...v_2^{\mu_k}v_1^{\mu_{k+1}}...v_1^{\mu_{2j-1}})$.  The
same sort of manipulations used in Eq.41 in section 3 can be used to show
that the matrix element, Eq.54, can be brought to the form, Eq.29.  In
this case however, the space parts of the invariant form factor are not
zero, indicating there are magnetic as well as electric form factors. By
working out several examples it can be seen that the magnetic form
factors have the general structure given in Eq.26.  The invariant form
factor is gotten from Eq.54 by evaluating the matrix element in the
standard (Breit) frame:
\begin{eqnarray}
F_\mu^a(Q^2)&=&D^j(B(v_1(st)))\bar{\sigma}_k(1,2)_\mu D^j(B(v_2(st)))
\nonumber\\
&&+\bar{D}^j(B(v_1(st)))\sigma_k(1,2)_\mu \bar{D}^j(B(v_2(st)))\\
F^a_{\mu=0}(Q^2)&=&\bar{\sigma}_k(1,2)_0 +\sigma_k(1,2)_0\\
F^a_{\mu=3}(Q^2)&=&0.
\end{eqnarray}
For $j=1/2$ this gives
\begin{eqnarray}
F^a_\mu(Q^2)&=&D^{1/2}(B(v_1(st)))\bar{\sigma}_\mu D^{1/2}(B(v_2(st)))
\nonumber\\
&&+\bar{D}^{1/2}(B(v_1(st)))\sigma_\mu\bar{D}^{1/2}(B(v_2(st)))\nonumber\\
F^a_{\mu=0}(Q^2)&=&\bar{\sigma}_0+\sigma_0\nonumber\\
&=&2I;\\
F^a_{\mu=i}(Q^2)&=&2i(\hat{z}\times\vec{\sigma})_i sh\alpha.
\end{eqnarray}
For $j=1$ there still is only one form factor, of the form
\begin{eqnarray}
F_\mu^a(Q^2)&=&(D^1(B(v_1(st)))\bar{\sigma}_{\mu\nu}D^1(B(v_2(st)))\nonumber\\
&&+\bar{D}^1(B(v_1(st)))\sigma_{\mu\nu}\bar{D}^1(B(v_2(st)))(v_1^{\nu}
(st)+v_2^{\nu}(st))\nonumber\\
&=&(D^1(B(v_1(st)))\bar{\sigma}_{\mu 0} D^1(B(v_2(st)))\nonumber\\
&&+\bar{D}^1(B(v_1(st)))\sigma_{\mu
0}\bar{D}^1(B(v_2(st))))2ch\alpha\nonumber\\ F^a_{\mu=0}(Q^2)&=&4ch\alpha
I;\\ F^a_{\mu=i}(Q^2)&=&4i(\hat{z}\times\vec{S})_i sh\alpha ch\alpha.
\end{eqnarray}
Starting with $j=3/2$ there will be several possibilities, depending on
the value of k;  here only the $j=3/2$ form factors are given:
\begin{eqnarray}F^a_\mu(Q^2)&=&(D^{3/2}(B(v_1(st)))
\bar{\sigma}_{\mu\mu_1\mu_2}D^{3/2}(B(v_2(st)))\nonumber\\
&&+\bar{D}^{3/2}(B(v_1(st)))\sigma_{\mu\mu_1\mu_2}\bar{D}^{3/2}(B(v_2(st)))
)(v_1^{\mu_1}v_1^{\mu_2}+v_2^{\mu_1}v_2^{\mu_2})\nonumber\\
F^a_{\mu=0}(Q^2)&=&4ch^2\alpha I+4\sigma_{033} sh^2\alpha\nonumber\\
&=&4 I+8sh^2\alpha \left[\begin{array}{cccc}1&&&\\&1/3&&\\&&1/3&\\
&&&1\end{array}\right]\\
F^a_{\mu=i}(Q^2)&=&(D^{3/2}(B(v_1(st)))\bar{\sigma}_{i00}
D^{3/2}(B(v_2(st))\nonumber\\
&&+\bar{D}^{3/2}(B(v_1(st)))\sigma_{i00}
\bar{D}^{3/2}(B(v_2(st))))2ch^2\alpha\nonumber\\
&&+(D^{3/2}(B(v_1(st)))\bar{\sigma}_{i33}D^{3/2}(B(v_2(st)))\nonumber\\
&&+\bar{D}^{3/2}(B(v_1(st)))\sigma_{i33}
\bar{D}^{3/2}(B(v_2(st))))2sh^2\alpha\nonumber\\
&=&4i(\hat{z}\times\vec{S})_i
sh\alpha+\frac{8}{9}i[(\hat{z}\times\vec{S})_i+2S_z
(\hat{z}\times\vec{S})_iS_z\nonumber\\
&&+2S_z^2(\hat{z}\times\vec{S})_i+2(\hat{z}\times\vec{S})_iS_z^2]
sh^3\alpha,
\end{eqnarray}
for $k=2$.  For $k=1$ there are similar expressions for the electric and
magnetic form factors;  because $v_1^{\mu_1}v_2^{\mu_2}$ replaces
$v_1^{\mu_1} v_1^{\mu_2}+v_2^{\mu_1}v_2^{\mu_2}$, the factor $8$ in Eq.62
is replaced by $-3$ and the factor 8 in Eq.63 by -4.

As seen from the general expression for the form factors, Eq.55, the
number of different form factors for integer spin is $j$ while for half
integer spin is $j+1/2$.  Combining these results with the elecric form
factors of section 3 then gives the correct number of total form factors,
namely
$2j+1$.
\section{Local Currents with Arbitrary Form Factors}
In the two previous sections currents were constructed from spin $j$
fields by differentiating the bilinear fields sufficiently many times to
contract with the gamma tensors to form a four-vector operator.  In this
section more derivatives will act on the fields to produce arbitrary
scalar form factors, while maintaining the locality of the current
operator.  The idea is best illustrated by starting with a scalar field,
for which the current is given before Eq.37;  such a current will not have
any form factors, and the functions $F_\mu^a$ and $F_\mu^c$ are given by
$\frac{(p_1+p_2)_\mu}{2mch\alpha}$ and 
$\frac{(p_1-p_2)_\mu}{2msh\alpha}$ respectively, in Eq.22.  By taking the
$n^{th}$ derivative of the fields, the following current results:
\begin{eqnarray}
J_\mu^{(n)}(x)&=&i(\frac{\partial}{\partial
x^{\nu_1}}...\frac{\partial}{\partial x^{\nu_n}})\phi^{\dagger}(x)
(\frac{\partial}{\partial x_{\nu_1}}...\frac{\partial}{\partial
x_{\nu_n}})
\frac{\partial}{\partial x^{\mu}}\phi(x)
\end{eqnarray}
minus a similar expression with the derivative of the free index,
$\frac{\partial}{\partial x^{\mu}}$ acting on the adjoint of the field,
needed for charge conservation.  If the fields are expanded in creation
and annihilation operators, the current becomes
\begin{eqnarray}
J_\mu^{(n)}(0)&=&\int dv_1dv_2[(p_1+p_2)_\mu(a^{\dagger}(p_1)a(p_2)
-b^{\dagger}(p_1)b(p_2))(v_1\cdot
v_2)^n\nonumber\\&&+(p_1-p_2)_\mu(a^{\dagger}(p_1)b^{\dagger}(p_2)
+a(p_1)b(p_2))(-v_1\cdot v_2)^n],
\end{eqnarray}
from which it is clear that the matrix elements are 
$F_\mu^a=(p_1+p_2)_\mu(v_1\cdot v_2)^n$ and 
$F_\mu^c=(p_1-p_2)_\mu(-v_1\cdot v_2)^n$.  If now each such current is
multiplied by a constant and summed over $n$, these matrix elements become
\begin{eqnarray}
F_\mu^a(1,2)&=&(p_1+p_2)_\mu f(v_1\cdot v_2)\\
F_\mu^c(1,2)&=&(p_1-p_2)_\mu f(-v_1\cdot v_2),
\end{eqnarray}
where $f(z)=\sum c_n z^n$ defines the power series for the function $f$
in terms of real coefficients, the first of which gives the charge. 
Notice that\\ $v_1\cdot 
v_2=ch^2\alpha+sh^2\alpha=1+2sh^2\alpha=1+\frac{Q^2}{2m^2}$, and shows
that
$f$ depends only on the momentum transfer squared.  If the
expressions in Eqs.66,67 are substituted into Eqs.34,35,36 expressing
locality, it should be possible to check the circumstances under which the
current remains local.  What is of particular interest is whether the current
remains local when the power seriers defines an analytic function.  This
question will be taken up in a later work.

It is now possible to combine these results with the results of the
previous two sections to obtain the most general current matrix elements
for a particle of spin $j$:
\begin{eqnarray}
F_\mu^a(1,2)&=&(p_1+p_2)_\mu\sum_{k}
f^E(v_1\cdot v_2)_k[D^j(B(v_1))\bar{\sigma}_k(1,2)D^j(B(v_2))\nonumber\\
&&+\bar{D}^j(B(v_1))\sigma_k(1,2)\bar{D}^j)](B(v_2))],\\
F_\mu^a(1,2)&=&\sum_{k} f^M(v_1\cdot
v_2)_k[D^j(B(v_1))\bar{\sigma}_k(1,2)_\mu
 D^j(B(v_2))\nonumber\\
&&+\bar{D}^j(B(v_1))\sigma_k(1,2)_\mu \bar{D}^j(B(v_2))],
\end{eqnarray}
where the first matrix element, Eq.68, corresponds to the most general
electric form factors of section 3, while the second matrix element
corresponds to the most general electric and magnetic form factor of
section 4.  Associated with these matrix elements are the corresponding
$F_\mu^c$ type matrix elements, which together form the most general local
current operator.  The factors $(-1)^k$ have been ignored
in these expressions, since they can all be absorbed in the $f_k$
coefficients.
\section{Conclusion}
This paper has shown how to construct local current operators in terms of
creation and annihilation operators for particles (and antiparticles) of
mass $m$ and spin $j$.  Since the construction of such current operators
involves arbitrary form factors, it is possible to view such particles
either as fundamental, or as composites of more fundamental constituents.

These results have been obtained by thinking of current operators as
elements of representations of the Lie algebra of the $U(N,N)$ type groups,
where the basis elements are bilinears in the creation and annihilation
operators.  In such a description the charge conjugation operation is an
automorphism acting on the group elements (or Lie algebra elements).  The
coefficients (one particle matrix elements) that multiply the bilinears are
partially constrained by requiring that the current operators have the
correct transformation properties under the extended Poincar\'{e} group, and
be conserved, hermitian and invariant under charge conjugation.  After these
constraints have been satisfied there are two remaining matrix elements
(see Eq.22) that themselves can be written as covariants times invariants
(the Wigner-Eckard theorem for the Poincar\'{e} group, see reference
\cite{f}).  The invariants can be further decomposed into form factors for
the various moments (charge, magnetic moment, quadrupole moment etc.) times
the matrix tensors introduced by Weinberg \cite{a}.  The representation given
for electric moments (Eq.25) and magnetic moments (Eq.26) generalizes well
known results for spin 1/2 and 1 to arbitrary spin particles.

The requirement that the current operator be local involves further
relations between the two remaining matrix elements.  Since it is not clear
how these relations might be satisfied (see Eqs.34, 35, 36), local fields
for arbitrary spin were introduced, following work of Weinberg \cite{a}. 
Bilinears in these local fields give current operators, the matrix elements
of which will satisfy-by construction-the locality requirements.

For a spin $j$ particle, there are $2j+1$ different possible currents that
can be built out of such local fields.  They break into two classes,
generalizing on the one hand currents made out of scalar particles, which
have only electric moments (there are $j+1$ such possibilities for integer
spin and $j+1/2$ for half-integer spin), while generalizing on the other
hand currents for spin 1/2 of the form $\bar{\Psi}\gamma_{\mu}\Psi$ (there
are
$j$ such possibilities for integer spin and $j+1/2$ for half-integer spin).  
For each of these classes the one particle matrix elements were given, in
section 3 for the scalar field generalization (see Eq.40) and in section 4
for the spin 1/2 generalization (see Eq.54).  Examples of these matrix
elements were also given for low spin values (see Eqs.47 to 51,and 58 to 63)
and it was also shown that the form of these matrix elements agrees with that
given by the more general analysis given in section 2 (see Eqs.25,26).

The values of the Breit frame matrix elements at zero momentum transfer
give the static moments of the particle.  If a definite current
is chosen the corresponding matrix element then fixes all the $2j+1$ static
moments of the particle;  for example, for a spin 1/2 particle, if the
current is chosen to be of $\gamma_{\mu}$ type, then the matrix element is
given in  Eqs.58,59  and the magnetic moment has a g factor of 2, as is
usually obtained from the Dirac equation.  Similarly, if the matrix elements
for a spin 1 particle are given by Eqs.60,61   then the magnetic moment and
quadrupole moment of the particle are fixed by the value of the matrix
element at zero momentum transfer.  But conversely, since there are 
$2j+1$ different possible currents, and each of them can be multiplied by an
arbitrary constant, using all the spin $j$ currents results in arbitrary
values for all the static moments.  These results are similar to analyses of
higher spin wave equations \cite{d}, where the choice of some wave equation
also implies that the static moments of the particle are fixed.  But
unlike wave equations, where additional constraints are imposed on the wave
function solutions of the wave equations, extracting the static properties
from matrix elements of current operators does not involve any constraints. 
As is well known these additional constraints imposed on wave functions lead
to acausal properties for higher spin particles
\cite{d}.  Further work is required to see what happens when the currents for
higher spin particles given in this paper are coupled to an external
electromagnetic field via the interaction given in Eq.11.

Aside from the current matrix elements extracted from the bilinears in free
local fields given in sections 3 and 4, there is also the possibility of
including arbitrary form factors while keeping the current operators local. 
This possibility was discussed in section 5 by letting the number of
derivatives acting on a current made out of fields be arbitrary.  By
allowing arbitrary constants in front of each power of derivatives, and then
adding up the resulting current operator that results, locality is
preserved.  The question that is raised in such a procedure is if
the number of derivatives go to infinity, what sorts of functions (if any)
keep the currents local.

 Assuming there are classes of functions for which
locality is preserved, the current operator given in Eq.22 will be a local
operator for this class of form factors and can be coupled to the photon
field to give the electromagnetic four-momentum operator for arbitrary spin
particles given in Eq.11.  Since the current operator is required to
transform as a four-vector under Lorentz transformations, it is also
necessary that the photon field transform as a four vector, in order that the
product transform as a scalar density.  Such a photon field can be
constructed as an induced representation of the Poincar\'{e} group, as
discussed in the next paper in this series \cite{c}.  For infinite degree of
freedom systems, it is crucial that both the current and photon operators be
local operators, in order that the point form equations, Eqs.1,2   be
satisfied.  On the other hand, for finite degree of freedom systems, locality
is not so important, and the general representation for current operators
given in Eq.22 for arbitrary form factors can be used in the electromagnetic
vertex to form the electromagnetic mass operator.  In both cases the current
operators for arbitrary spin particles can be used to produce an
electromagnetic interaction in point form relativistic quantum mechanics.
\section{Appendix:  Finite Dimensional Representations of the Lorentz
Group and Free Local Fields}
Finite dimensional representations of the Lorentz group play a crucial
role in the analysis of currents.  Let $\Lambda$ denote an element of the
(proper) Lorentz group, SO(1,3), with $\Lambda g \Lambda^T=g, g=diag
(1,-1,-1,-1)$ and define the column vector $x:=(t,\vec{x})$;  then a
Lorentz transformation $\Lambda$ sends $x$ to $x^{'}=\Lambda x$.  This
can also be written in index notation as $x^{\mu}->x^{'\mu}=\Lambda^\mu
_\nu x^{\nu}$.  The Lorentz invariant is  $x\cdot x=x^Tgx=x^{\mu}
g_{\mu\nu}x^{\nu}$.  Both the matrix and index notation will be used in
this paper, depending on the context.

Any element of SO(1,3) can be decomposed into a boost 
times a rotation, $\Lambda=B(v) R$, where $B(v)$ is a boost, a coset
representative of SO(1,3) with respect to the (proper) rotation group 
SO(3).  In this paper $B(v)$ will always be taken to be a canonical spin
boost, namely
\begin{eqnarray}
B(v)&=&R(\hat{v})\Lambda_z(\alpha)R^{-1}(\hat{v})\\
&=&\left[\begin{array}{cc}v_0&\vec{v}^T\\
\vec{v}&{I+\frac{\vec{v}\otimes\vec{v}^T}{v_0+1}}\end{array}\right]\\
R(\hat{v})&=&R_z(\phi)R_y(\theta)\\
\Lambda_z(\alpha)&=&\left[\begin{array}{cccc}ch\alpha&0&0&sh\alpha\\
0&1&0&0\\
0&0&1&0\\
sh\alpha&0&0&ch\alpha\end{array}\right],\
\end{eqnarray}
with $\phi$ the azimuthal angle of the z axis rotation $R_z(\phi)$ and 
$\theta$ the polar angle of the y axis rotation $R_y(\theta)$ forming the
unit vector $\hat{v}$.  $v$ is the four-velocity satisfying $v\cdot v=1$ 
and
$ch\alpha=v^0=v_0, sh\alpha=|\vec{v}|$.  In particular  $B(v)p^{rest}=
p=mv$, with $p^{rest}=(m,\vec{0})$.

A Lorentz transformation followed by a boost gives a Wigner rotation:
\begin{eqnarray}
\Lambda B(v)&=&B(\Lambda v) R_W(v,\Lambda)\\
R_W(v,\Lambda):&=&B^{-1}(\Lambda v) \Lambda B(v);\\
R_W(v,R)&=&R,\
\end{eqnarray}
where in Eq.76 use has been made of the property of canonical spin boosts
that the Wigner rotation of a rotation is that rotation.

In this paper only those Lorentz representations commonly denoted by
$(j,0)$ and $(0,j)$ will be needed, rather than the more general
$(j,j^{'})$ representations.  The representations $(j,0)$ are the
so-called holomorphic representations of GL(2,C) and can be obtained as
induced representations on Bargmann space \cite{e}.  From this analysis
only the matrix elements
\begin{eqnarray}
D^{(j,0)}(\Lambda_z(\alpha))&=&diag(e^{j\alpha}...e^{-j\alpha})\\
D^{(j,0)}(R)&=&D^j(R)
\end{eqnarray}
will be needed.  Combining the coset decomposition of a Lorentz
transformation with the expression for the canonical spin boost, Eq.70,
then gives the double coset decomposition of a Lorentz transformation,
$\Lambda=R\Lambda_z(\alpha)R^{'}$, so that the representation of a
Lorentz transformation is given by
$D^j(\Lambda)=D^j(R)D^j(\Lambda_z(\alpha))D^j(R^{'})$ where $j$ means 
$(j,0)$.

From the definition of canonical spin boost, Eq.70, it follows that
\begin{eqnarray}
D^j(B(v))&=&D^j(R(\hat{v}))D^j(\Lambda_z(\alpha))D^j(R^{-1}(\hat{v}))\\
&=&D^{j}(B(v))^{\dagger};\\
D^{1/2}(B(v))&=&\frac{1}{\sqrt{2(v_0+1)}}\left[\begin{array}{cc}v_+ +1
&v_{\perp}^{\ast}\\
v_{\perp}&v_- +1\end{array}\right]
\end{eqnarray}
with $v_{\pm}=v_0\pm v_3$.

The generalized $\sigma$ matrices play an important role in the analysis
of currents.  They are defined by
\begin{eqnarray}
\sigma_{{\mu_1}...{\mu_{2j}}}v^{\mu_1}...v^{\mu_{2j}}:&=&D^j(B(v)B(v))
\end{eqnarray}
and are symmetric (in all indices), traceless, matrix tensors; 
because of our different metric and sign conventions, we have chosen to
denote them by $\sigma$ (or $
\bar{\sigma})$, rather than $t$ (or $\bar{t}$) as originally defined by
Weinberg \cite{a} in his Eq.A10 (Eq.A38).  In spin indices the sigmas are
$2j+1$ by $2j+1$ matrices, which is also the number of traceless symmetric
tensors.

From the definition given in Eq.82 it follows that
\begin{eqnarray}
D^j(\Lambda)\sigma_{\mu_{1}...\mu_{2j}}v^{\mu_{1}}...v^{\mu_{2j}}D^j
(\Lambda)^{\dagger}
&=&D^j(\Lambda)D^j(B(v)B(v))D^{j}(\Lambda)^{\dagger}\nonumber\\
&=&D^j(\Lambda B(v))D^{j}(\Lambda B(v))^{\dagger}\nonumber\\
&=&D^j(B(\Lambda v)R_W)D^{j}(B(\Lambda v)R_W)^{\dagger}\nonumber\\
&=&D^j(B(\Lambda v)B(\Lambda v))\nonumber\\
&=&\sigma_{\mu_{1}...\mu_{2j}}(\Lambda v)^{\mu_{1}}...(\Lambda
v)^{\mu_{2j}},
\end{eqnarray}
which implies that $\sigma$ transforms as a $2j^{th}$ rank tensor under
Lorentz transformations.  In particular, from Eq.81
\begin{eqnarray}
D^{1/2}(B(v)B(v))&=&\left[\begin{array}{cc}v_+&v_{\perp}^{\ast}\\
v_{\perp}&v_-\end{array}\right]\nonumber\\
&=&\sigma_{\mu}v^{\mu};\\
D^{1/2}(\Lambda)\sigma_{\mu}v^{\mu}D^{1/2}(\Lambda)^{\dagger}&=&
\sigma_{\mu}(\Lambda v)^{\mu}\nonumber\\
&=&\sigma_{\mu}\Lambda^{\mu}_{\nu}v^{\nu}.
\end{eqnarray}

The conjugate representation $\bar{D}^j(\Lambda)$, related to $D^{(0,j)}$,
is defined by $\bar{D}^j(\Lambda):=D^{j}(\Lambda^{-1})^{\dagger}$; 
corresponding to these representations are the conjugate sigma matrix
tensors defined by
\begin{eqnarray}
\bar{\sigma}_{\mu_{1}...\mu_{2j}}v^{\mu_1}...v^{\mu_{2j}}:&=&
\bar{D}^j(B(v)B(v));\\
\bar{D}^j(\Lambda)\bar{\sigma}_{\mu_{1}...\mu_{2j}}v^{\mu_{1}}...v^{\mu_{2j}}
\bar{D}^{j}(\Lambda)^{\dagger}&=&\bar{\sigma}_{\mu_{1}...\mu_{2j}}(\Lambda
v)^{\mu_1} ...(\Lambda v)^{\mu_{2j}}.
\end{eqnarray}

The sigma tensors can be given explicitly by making use of their
symmetric traceless properties;  in particular, for $v=(ch\alpha,0,0,
sh\alpha)$ it is possible to write
\begin{eqnarray}
D^j(B(v)B(v))&=&diag(e^{2j\alpha}...e^{-2j\alpha})\nonumber\\
&=&\sigma_{0...0}(ch\alpha)^{2j}+2j\sigma_{30...0}(ch\alpha)^{2j-1}sh\alpha
\nonumber\\
&&+\frac{2j(2j-1)}{2}\sigma_{330...0}(ch\alpha)^{2j-2}(sh\alpha)^2+...;\\
\bar{D}^j(B(v)B(v))&=&diag(e^{-2j\alpha}...e^{2j\alpha})\nonumber\\
&=&\bar{\sigma}_{0...0}(ch\alpha)^{2j}+2j\bar{\sigma}_{30...0}(ch\alpha)^{2j-1}
sh\alpha+...
\end{eqnarray}
By differentiating Eq.88 sufficiently many times with respect to 
$\alpha$ and then setting $\alpha$ to zero, it follows that
\begin{eqnarray}
\sigma_{0...0}&=&I\\
\sigma_{30...0}&=&\frac{1}{j}S_z\\
\sigma_{330...0}&=&\frac{2}{j(2j-1)}S_z^2-\frac{1}{2j-1}I
\end{eqnarray}
with similar expressions for the higher order sigma matrices.  The $S_i$
are  spin angular momentum matrices and
$I$ is the identity matrix (of dimension $2j+1$). Combining the results
for the special values of the sigma matrices given in Eqs.91,92 with the
general form for symmetric traceless tensors then results in the
following lower order expressions for the sigma matrices:
\begin{eqnarray}
\sigma_{i0...0}&=&\frac{1}{j}S_i\\
\sigma_{i_1i_2
0...0}&=&\frac{1}{j(2j-1)}(S_{i_1}S_{i_2}+S_{i_2}S_{i_1})-\frac{1}{2j-1}
\delta_{i_1i_2}I
\end{eqnarray}

Also, differentiating
Eq.88 with respect to alpha $n$ times gives factors of
$(2S_z)^n$, while differentiating Eq.89 with respect to alpha $n$ times
gives factors of
$(-2S_z)^n$.  Thus the two sigma tensors are related to one another by
\begin{eqnarray}
\bar{\sigma}_{i_{1}...i_{n}0...0}&=&\sigma_{i_{1}...i_{n}0...0},
n\;even\\
\bar{\sigma}_{i_{1}...i_{n}0...0}&=&-\sigma_{i_{1}...i_{n}0...0}, n\:odd
\end{eqnarray}

Finally, to construct the local fields needed for making current
operators, it is necessary to work out the properties of the discrete
transformations, especially parity.  The parity operation is defined by 
$Px=(x_0,-\vec{x})=gx$.  Further, on states,
$U_P|p\sigma>=\eta|gp,\sigma>$, where $\eta$ is the intrinsic parity. 
Also, from Eq.70
\begin{eqnarray}
B^{-1}(\vec{v})&=&R(\hat{v})\Lambda_z^{-1}(\alpha)R^{-1}(\hat{v})\nonumber\\
&=&R(\hat{v})R_x(\pi)\Lambda_z(\alpha)R_x(\pi)R^{-1}(\hat{v})\nonumber\\
&=&R(-\hat{v})\Lambda_z(\alpha)R^{-1}(-\hat{v})\nonumber\\
&=&B(-\vec{v}).\\
D^j(B(-\vec{v}))&=&D^j(B^{-1}(\vec{v}))\nonumber\\
&=&\bar{D}^j(B(\vec{v})),
\end{eqnarray}
a result that is needed for the transformation properties of the fields
under parity.

Following Weinberg \cite{a} local spin $j$ fields are defined by
\begin{eqnarray}
\phi_n(x)&=&\sum \int dv D_{n\sigma}^j(B(v))[a(p\sigma)e^{-ip\cdot
x}+C^{-1}_{\sigma\sigma^{'}}b^{\dagger}(p\sigma^{'})e^{ip\cdot
x}]\\ U_a\phi_n(x)U_a^{-1}&=&\phi_n(x+a)\\
U_\Lambda\phi_n(x)U_\Lambda^{-1}&=&\sum
D_{nn^{'}}^j(\Lambda^{-1})\phi_{n^{'}} (\Lambda x),
\end{eqnarray}
with locality properties given by
\begin{eqnarray}
{[}\phi_n(x),\phi^{\dagger}_{n^{'}}(0)]_{\pm}&=&\sum \int dv D^j_{nn^{'}}
(B(v)B(v))[e^{-ip\cdot x}{\pm}e^{ip\cdot x}]\\
&=&\frac{1}{m^2j}\sigma_{\mu_{1}...\mu_{2j}}(i\frac{\partial}
{\partial_{\mu_1}}...i\frac{\partial}{\partial_{\mu_{2j}}})\int
dv[e^{-ip\cdot x}{\pm}(-1)^{2j}e^{ip\cdot x}],\nonumber
\end{eqnarray}
where the measure $dv=\frac{d^3 v}{2 v_0}$ and $C$ is the
conjugation matrix defined by $D^j(R)^{\ast}=CD^j(R)C^{-1}$.  As shown by
Weinberg such fields will be local only if ${\pm}(-1)^{2j}=-1$, which
gives the usual connection between spin and statistics.

Under parity the $2j+1$ component field $\phi$ transforms into a second
$2j+1$ component field $\chi$:
\begin{eqnarray}
U_P\phi_n(x)U_P^{-1}&=&\sum \int dv D_{n\sigma}^j(B(v))[\eta_P
a(gp,\sigma)e^{-ip\cdot x}\nonumber\\
&&+\bar{\eta}_P
C^{-1}_{\sigma\sigma^{'}}b^{\dagger}(gp,\sigma^{'})e^{ip\cdot x}]\\
 &=&\sum \int dv
D_{n\sigma}^j(B^{-1}(v))^{\dagger}[\eta_Pa(p\sigma)e^{-ip\cdot
gx}\nonumber\\
&&+\bar{\eta}_PC^{-1}_{\sigma\sigma^{'}}
b^{\dagger}(p\sigma^{'})e^{ip\cdot gx}]\\ &=&\eta\chi_n(gx);\\
\chi_n(x):&=&\sum \int dv\bar{D}^j(B(v))[a(p\sigma)e^{-ip\cdot
x}\nonumber\\ &&+(-1)^{2j}C^{-1}_{\sigma\sigma^{'}}
b^{\dagger}(p\sigma^{'})e^{ip\cdot x}],
\end{eqnarray}
with the intrinsic parity of the antiparticle related to the intrinsic
parity of the particle by $\eta\bar{\eta}=(-1^{2j})$. 

The two fields are related by the sigma tensors in the following way:
\begin{eqnarray}
\bar{\sigma}_{\mu_1...\mu_{2j}}(i\frac{\partial}{\partial x_{\mu_1}}...i
\frac{\partial}{\partial x_{\mu_{2j}}})\phi(x)&=&m^{2j}\sum\int dvD^j(B^{-1}
(v)B^{-1}(v)B(v))\nonumber\\
&&[a(p\sigma)e^{-ip\cdot
x}+(-1)^{2j}C^{-1}_{\sigma\sigma^{'}}b^{\dagger}(p\sigma^{'})e^{ip\cdot
x}]\nonumber\\
&=&m^{2j}\chi(x);\\
\sigma_{\mu_1...\mu_{2j}}(i\frac{\partial}{\partial x_{\mu_1}}
...i\frac{\partial}{\partial x_{\mu_{2j}}})\chi(x)&=&m^{2j}\phi(x).
\end{eqnarray}

By combining the two fields into a $2(2j+1)$ component field and forming
a generalized $\gamma$ matrix from the two sigma matrices, a generalized
Dirac equation can be written as
\begin{eqnarray}
(-i\gamma_{\mu_1...\mu_{2j}}(\frac{\partial}{\partial x_{\mu_1}}...
\frac{\partial}{\partial x_{\mu_{2j}}})+m^{2j})\Psi(x)&=&0,
\end{eqnarray}
where
\begin{eqnarray}
\Psi(x):&=&\left[\begin{array}{c}\phi(x)\\
\chi(x)\end{array}\right],\\
\gamma_{\mu_1...\mu_{2j}}:&=&-i(i^{2j})\left[\begin{array}{cc}0&\sigma_{\mu_1...\mu_{2j}}\\
\bar{\sigma}_{\mu_1...\mu_{2j}}&0\end{array}\right],
\end{eqnarray}
which agrees with the usual Dirac equation when $j=1/2$.  Also, under
parity, using Eq.103, $U_P\Psi(x)U_P^{-1}=\eta\beta\Psi(gx)$, where
$\beta=\left[\begin{array}{cc}0&I\\I&0\end{array}\right]$ is the
generalized $\beta$ matrix and $I$ the $(2j+1)$ dimensional identity
matrix.  The generalized Dirac equation, viewed as a wave equation presents
difficulties connected with minimal substitution and acausal solutions; for a
discussion of these points, see reference \cite{g}.

\end{document}